\documentstyle[11pt]{article}

\title{Supersymmetric grandunification and fermion masses}
\author{Borut Bajc\\
J. Stefan Institute, 1001 Ljubljana, Slovenia\\
borut.bajc@ijs.si}

\begin{document}
\maketitle

\begin{abstract}
A short review of the status of supersymmetric grand unified 
theories and their relation to the issue of fermion masses and 
mixings is given. 
\end{abstract}

\section{Why Grandunification?}
\label{whygut}

There are essentially three reasons for trying to build grand unified 
theories (GUTs) beyond the standard model (SM).

\begin{itemize}
\item why should strong, weak and electromagnetic couplings 
in the SM be so different despite all corresponding to 
gauge symmetries?
\item there are many disconnected matter representations in the SM 
(3 families of $L$, $e^c$, $Q$, $u^c$, $d^c$)
\item quantization of electric charge (in the SM model there are two 
possible explanations - anomaly cancellation and existence of magnetic 
monopoles - both are naturally embodied in GUTs)
\end{itemize}

\section{How to check a GUT?}
\label{checkgut}

I will present here a very short review of some generic features, 
predictions and drawbacks of GUTs. Details of some topics will be 
given in the next section.

\subsection{Gauge coupling unification}

This is of course a necessary condition for any GUT to work. 
As is well known, the SM field content plus the desert assumption 
do not lead to the unification of the three gauge couplings. 
However, the idea of low-energy supersymmetry (susy), 
i.e. the minimal supersymmetric standard model (MSSM) instead 
of the SM at around TeV and again the assumption of the desert gives 
a quite precise unification of gauge couplings at 
$M_{GUT}\approx 10^{16}$ GeV \cite{Dimopoulos:1981yj}. 
Clearly there is no a-priori reason for three functions to cross 
in one point, so this fact is a strong argument for supersymmetry. 
One gets two bonuses for free in this case. First, the hierarchy 
problem gets stabilized, although not really solved, since 
the famous doublet-triplet (DT) problem still remains. 
Secondly, at least in principle one can get an insight into the 
reasons for the electroweak symmetry breaking: why the Higgs 
(other bosons in MSSM) mass squared is negative (positive) at 
low energy \cite{Inoue:1982pi}.

\subsection{Fermion masses and mixings}

Although GUTs are not theories of flavour, they bring 
constraints on the possible Yukawas. In the MSSM the 
Yukawa sector is given by 

\begin{equation}
W_Y=HQ^TY_Uu^c+\overline{ H}Q^TY_Dd^c+\overline{ H}L^TY_Ee^c\;,
\end{equation}

\noindent
and the complex $3\times 3$ generation matrices 
$Y_{U,D,E}$ are arbitrary. 
However, in a GUT the matter fields $Q$, $L$, $u^c$, $d^c$, $e^c$ 
fields live together in bigger representations, so one expects 
relations between quark and lepton Yukawa matrices. 

Take for example the SO(10) GUT. All the MSSM matter fields of each 
generation live in the same representation, the $16$ 
dimensional spinor representation, which contains and thus 
predicts also the right-handed neutrino. At the same time the minimal 
Higgs representation, the $10$ dimensional representation contains 
both doublets $H$ and $\overline{ H}$ of the MSSM (plus one color triplet 
and one antitriplet). The only renormalizable SO(10) invariant one can 
write down is thus 

\begin{equation}
W_Y=10_H16Y_{10}16\;,
\end{equation}

\noindent
which is however too restrictive, since it gives on top of the well 
satisfied (for large $\tan{\beta}$) relation $y_b=y_\tau=y_t$ for the 
third generation, the much worse predictions for the first two 
generations ($y_s=y_\mu=y_c$ and $y_d=y_e=y_u$) and no mixing 
($\theta_c=0$) at all. 

How to improve the fit? Let us mention two possibilities:

(1) Introduce new Higgs representations: although another $10_H$ can 
help with the mixing, the experimentally wrong relations $m_d=m_e$ 
and $m_s=m_\mu$ still occur, because the two bi-doublets in the two 
$10_H$ leave invariant the Pati-Salam SO(6)=SU(4)$_C$, so the leptons 
and quarks are still treated on the same footing. 
So the idea to pursue is to 
introduce bidoublets which transform nontrivially under the Pati-Salam 
SU(4) color. This can be done for example by 
introducing a $\overline{ 126}_H$, 
which couples to matter as $\Delta W_Y=\overline{ 126}_H16Y_{126}16$ and 
which gets a nontrivial vev in the $(2,2,15)_H$ SU(3) color singlet 
direction \cite{Lazarides:1980nt,Babu:1992ia}. 

(2) Another possibility is to include the effects of nonrenormalizable 
operators. These operators can cure the problem and at 
the same time ease the proton decay 
constraints. The drawback is the loss of predictivity. 

\subsection{Proton decay}

This issue is connected to 

(1) R-parity. It is needed to avoid fast proton decay. At the 
nonrenormalizable level one could for example have terms leading to 
R-parity violation of the type $16^316_H/M_{Pl}$. For this reason it is 
preferable to use the $126_H$ representation instead of the $16_H$. 
It is possible to show that such a SO(10) with $126_H$ has an exact 
R-parity \cite{Aulakh:1997ba} at all energies 
without the introduction of further symmetries. 

(2) DT splitting problem: Higgs SU(2)$_L$ doublets and 
SU(3)$_C$ triplets live usually in the same GUT multiplet; but 
while the SU(2)$_L$ doublets are light ($\approx M_W\ll M_{GUT}$), 
the SU(3)$_C$ triplets should be very heavy ($\ge M_{GUT}$) to 
avoid a too fast proton decay. For example, the proton lifetime in 
susy is proportional to $M_T^2$, which can give a lower limit 
to the triplet mass \cite{Hisano:1992jj}, although this limit depends 
on the yet unknown supersymmetry breaking sector \cite{Bajc:2002bv}. 

The solutions to the DT problem depend on the gauge group considered, 
but in general models that solve it are not minimal and necessitate 
of additional Higgs sectors. For example the missing partner mechanism 
\cite{Masiero:1982fe} in SU(5) needs at least additional 
$75_H$, $50_H$ and $\overline{ 50}_H$ representations. The same is 
true for the missing vev mechanism in SO(10) 
\cite{Dimopoulos:xm}, where the $45_H$ 
and extra $10_H$ Higgses must be introduced. Also the nice idea of 
GIFT (Goldstones Instead of Fine Tuning) \cite{Berezhiani:1995sb} 
can be implemented only by complicated models, while discrete 
symmetries for this purpose can be used with success only in 
connection with non-simple gauge groups \cite{Barbieri:1994jq}. 
Of course, although not very natural, any GUT can "solve" 
the problem phenomenologically, i.e. simply fine-tuning 
the model parameters. 

Clearly, whatever is the solution to the DT problem, the proton 
lifetime depends in a crucial (powerlike) way on the triplet mass. 
And this mass can be difficult to determine from the gauge coupling 
unification condition even in specific models 
because of the unknown model parameters 
\cite{Bachas:1995yt} or use of high representations 
\cite{Dixit:1989ff}. On top of this there can be large uncertainties 
in the triplet Yukawa couplings \cite{Dvali:1992hc}. 
All this, together with the phenomenologically completely unknown soft 
susy breaking sector, makes unfortunately proton decay not a very neat 
probe of supersymmetric grandunification \cite{Bajc:2002bv}.
Of course, if for some reason the DT mechanism is so efficient to make 
the $d=6$ operators dominant (for a recent analysis in some 
string-inspired models see \cite{Friedmann:2002ty}), 
then the situation could be simpler to analyse \cite{Mohapatra:yj}, 
although many uncertainties due to fermion mixing matrices still exist in 
realistic nonminimal models \cite{Nandi:1982ew}. 
Unfortunately there is little hope to detect proton decay in this case, 
unless $M_{GUT}$ is lower than usual \cite{Aulakh:2000sn}. 

\subsection{Magnetic monopoles}

Since magnetic monopoles 
are too heavy to be produced in colliders, the 
only hope is to find them as relics from the cosmological GUT phase 
transitions. Their density however strongly depends on the cosmological 
model considered. Unfortunately, the Rubakov-Callan effect 
\cite{Rubakov:fp} leads to the non observability 
of GUT monopoles, at least in any foreseeable future. Namely, these 
monopoles are captured by neutron stars and the resulting astrophysical 
analyses \cite{Freese:1983hz} limits the monopole flux at earth twelve 
orders of magnitude below the MACRO limit \cite{Giacomelli:2003yu}. 

This is very different from the situation in the Pati-Salam (PS) 
theory. Even in the minimal version the PS scale can be much lower 
than the GUT scale \cite{Melfo:2003xi}, as low as $10^{10}$ GeV. 
the resulting monopoles are then too light to be captured by 
neutron stars and their flux is not limited due to the Rubakov-Callan 
effect. Furthermore, MACRO results are not applicable for such light 
monopoles \cite{Giacomelli:2003yu}. 

\subsection{Low energy tests}

There are many different possible tests at low-energy, like for example 
the flavour changing neutral currents (see for example 
\cite{Barbieri:1994pv}) 
or the electric dipole moments \cite{Masina:2003iz}. In the latter 
case the exact value of the triplet mass is much less important than in 
proton decay, but the uncertainties due to the susy breaking sector 
are still present. In some of these tests like neutron-antineutron 
oscillation we can get positive signatures only for specific 
models due to very high dimensional 
operators involved \cite{Babu:2001qr}. 

\section{Fermion masses and mixings}

The regular pattern of 3 generations suggests some sort of flavour 
symmetry. 

One way, and the most ambitious one, is to consider the flavour 
symmetry group as part (subgroup) of the grand unified gauge 
symmetry (described by a simple group). In such an approach all 
three generations come from the same GUT multiplet. For example, 
in SU(8) the 216 dimensional representation gets 
decomposed under its SU(5) subgroup into three copies (generations) 
of $\overline{ 5}$ and $10$ with additional SU(5) multiplets. 
Similarly, in the SO(18) GUT, the 256 dimensional spinorial 
representation is nothing else than 8 generations of 
$(16+\overline{ 16})$ in the SO(10) language. The problem in all 
these theories is what to do with all the extra light particles 
\cite{Wilczek:1981iz}. 

Another possibility is to consider the product of the flavour 
(or, in general, extra) symmetry with the GUT symmetry (simple) 
group. In the context of SO(10) GUTs most of them use small 
representations for the Higgses, like $16_H$, $\overline{ 16}_H$ 
and $45_H$. The philosphy is to consider all terms allowed by 
symmetry, also nonrenormalizable. The DT problem can be naturally 
solved by some version of the missing vev mechanism, which however 
means that many multiplets are usually needed. Such models are quite 
successfull \cite{Albright:2000sz}, although the 
assumed symmetries are a little bit ad-hoc. There is also a huge 
number of different models with almost arbitrary flavour symmetry 
group, but unfortunately there is no room to describe them here 
(see for example the recent review \cite{Chen:2003zv}).

What we will consider in the following is instead a SO(10) GUT 
with no extra symmetry at all. We want to see how far we can go 
with just the grand unified gauge symmetry alone. To ensure automatic 
R-parity, we are forced not to use the $16_H$ and $\overline{ 16}_H$ 
Higgses, but instead a pair of $126_H$ and $\overline{ 126}_H$ 
(5 index antisymmetric representations, one self-dual, 
the other anti-self-dual; both of them are needed in order not to break 
susy at a large scale). In fact under R-parity the bosons of 
$16$ are odd, while those of $126$ are even, since 

\begin{equation}
R=(-1)^{3(B-L)+2S}
\end{equation}

\noindent
\cite{Mohapatra:su}, and the relevant vev in the SU(5) singlet 
directions have $B-L=1$ for $16_H$ ($\nu^c$), while it has 
$B-L=2$ for $126$ (the mass of $\nu^c$). 

So the rules of the game are: stick to renormalizable operators only, 
consider SO(10) as the only symmetry of the model, take the minimal 
number of multiplets (it does not mean the minimal number of fields!) 
that is able to give the correct symmetry breaking pattern
SO(10)$\to$SU(3)$\times$SU(2)$\times$U(1). Such a theory is 
given by \cite{Clark:ai} (see however \cite{Aulakh:1982sw} 
for a similar approach): on top of the usual three 
generations of $16$ dimensional matter fields, it contains four 
Higgs representations: $10_H$, $126_H$, $\overline{ 126}_H$ and
$210_H$ (4 index antisymmetric). It has been shown recently 
\cite{Aulakh:2003kg} that this theory is also the minimal GUT, 
i.e. it has the minimal number of model parameters, being still 
perfectly realistic (not in contradiction with any experiment). 

As we have seen, the $\overline{ 126}_H$ multiplet is needed both 
to help the $10_H$ multiplet in fitting the fermion masses and mixings, 
and for giving the mass to the right-handed neutrino without explicitly 
breaking R-parity. Let us now see, why the $210_H$ representation 
is needed.

The Yukawa sector is given by 

\begin{equation}
\label{wy}
W_Y=10_H16Y_{10}16+\overline{ 126}_H16Y_{126}16\;.
\end{equation}

The fields decompose under the 
SU(2)$_L\times$SU(2)$_R\times$SU(4)$_C$ subgroup as 

\begin{eqnarray}
10_H&=&(2,2,1)+(1,1,6)\;,\\
16&=&(2,1,4)+(1,2,\overline{ 4})\;,\\
\overline{ 126}_H&=&(1,3,10)+(3,1,\overline{ 10})+(2,2,15)+(1,1,6)\;.
\end{eqnarray}

The right-handed neutrino $\nu^c$ lives in $(1,2,\overline{ 4})$ of $16$, 
so it can get a large mass only through the second term in (\ref{wy}):

\begin{equation}
\label{mnr}
M_{\nu_R}=\langle (1,3,10)_{\overline{126}}\rangle Y_{126}\;,
\end{equation}

\noindent
where $\langle (1,3,10)\rangle$ is the scale of the SU(2)$_R$ 
symmetry breaking $M_R$, which we assume to be large, 
${\cal O}(M_{GUT})$. 

In order to get realistic masses we need 

\begin{eqnarray}
\langle (2,2,1)_{10}\rangle&=&
\pmatrix{
   v_{10}^d
&  0
\cr   
   0
&  v_{10}^u
\cr   }
\ne 0\;,\\
\langle (2,2,15)_{\overline{126}}\rangle&=&
\pmatrix{
   v_{126}^d
&  0
\cr   
   0
&  v_{126}^u
\cr   }\ne 0\;,
\end{eqnarray}

\noindent
which contribute to the light fermion masses as

\begin{eqnarray}
\label{mu}
M_U&=&v_{10}^uY_{10}+v_{126}^uY_{126}\;,\\
\label{md}
M_D&=&v_{10}^dY_{10}+v_{126}^dY_{126}\;,\\
\label{mnd}
M_{\nu_D}&=&v_{10}^uY_{10}-3v_{126}^uY_{126}\;,\\
\label{me}
M_E&=&v_{10}^dY_{10}-3v_{126}^dY_{126}\;.
\end{eqnarray}

The factor of $-3$ for leptons in the contribution from 
$\overline{126}_H$ comes automatically from the fact that 
the SU(3)$_C$ singlet in the adjoint $15$ of SU(4)$_C$ is in the 
$B-L$ direction diag$(1,1,1,-3)$. This is clearly absent in 
the contribution from $10_H$, which is a singlet under the full 
SU(4)$_C$. 

The light neutrino mass comes through the famous see-saw 
mechanism \cite{Mohapatra:1979ia}. From 

\begin{equation}
W={1\over 2}\nu^{cT}M_{\nu_R}\nu^c+\nu^{cT}M_{\nu_D}\nu_L+...
\end{equation}

\noindent
one can integrate out the heavy right-handed neutrino $\nu^c$ 
obtaining the effective mass term for the light neutrino states 
$M_N=-M_{\nu_D}^TM_{\nu_R}^{-1}M_{\nu_D}$. As we will now see, there 
is another contribution in our minimal model. 

(1) We saw that both $\langle (2,2,1)_{10}\rangle$ and 
$\langle (2,2,15)_{126}\rangle$ need to be nonzero and obviously 
${\cal O}(M_W)$. With $10_H$, $126_H$ and $\overline{126}_H$ 
Higgses one can write only two renormalizable invariants:

\begin{equation}
\label{wh}
W_H={1\over 2}M_{10}10_H^2+M_{126}126_H\overline{126}_H\;,
\end{equation}

\noindent
where $M_{10},M_{126}\approx {\cal O}(M_{GUT})$ or larger due 
to proton decay constraints. So the mass term looks like 

\begin{equation}
{1\over 2}(10_H,126_H,\overline{126}_H)
\pmatrix{
   M_{10}
&  0
&  0
\cr 
   0  
&  0
&  M_{126}
\cr
   0
&  M_{126}
&  0
\cr   }
\pmatrix{
   10_H
\cr
   126_H
\cr
   \overline{126}_H
\cr   }\;.
\end{equation}

Clearly all the doublets have a large positive mass, so their 
vev must be zero. Even fine-tuning cannot solve the DT problem in 
this case. So the idea to overcome this obstacle is to mix in some way 
$10_H$ with $\overline{126}_H$ ($126_H$), and after that fine-tune 
to zero one combination of doublet masses. So the new mass matrix 
should look something like 

\begin{equation}
\label{mxy}
\pmatrix{
   M_{10}
&  x
&  y
\cr 
   x  
&  0
&  M_{126}
\cr
   y
&  M_{126}
&  0
\cr   }
\end{equation}

\noindent
with $x,y$ denoting such mixing. The light Higgs doublets will thus be 
linear combinations of the fields in $(2,2,1)_{10}$ and $(2,2,15)_{126_H,
\overline{126}_H}$ and this will get a nonzero vev after including the soft 
susy breaking masses. 

(2) The minimal representation that can mix $10$ and $\overline{126}$ 
is $210$, as can be seen from $10\times\overline{126}=210+1050$. 
$210$ is a 4 index antisymmetric SO(10) representation, which 
decomposes under the Pati-Salam subgroup as

\begin{equation}
210=(1,1,1)+(1,1,15)+(1,3,15)+(3,1,15)+
(2,2,6)+(2,2,10)+(2,2,\overline{10})\;.
\end{equation}

Of course one can now add other renormalizable terms to (\ref{wh}), 
and all such new terms are (in a symbolic notation)

\begin{equation}
\label{dwh}
\Delta W_H=210_H^3+210_H^2+210_H126_H\overline{126}_H+
210_H10_H126_H+210_H10_H\overline{126}_H\;.
\end{equation}

The last two terms are exactly the ones needed for the mixings 
between $10_H$ and $126_H$ ($\overline{126}_H$), i.e. contributions 
to $x,y$ in (\ref{mxy}). It is possible to show that $W_H+\Delta W_H$ 
are just enough for SO(10)$\to$SM. In the case of single-step breaking 
one thus has

\begin{eqnarray}
\langle (1,1,1)_{210}\rangle&\approx&
\langle (1,1,15)_{210}\rangle\approx
\langle (1,3,15)_{210}\rangle\approx\nonumber\\
\langle (1,3,\overline{10})_{126}\rangle&\approx&
\langle (1,3,10)_{\overline{126}}\rangle\approx
M_{GUT}\;.
\end{eqnarray}

(3) Now however there are five bidoublets that mix, 
since $(2,2,10)$ and $(2,2,\overline{10})$ from $210_H$ 
also contribute. To be honest, there is only one neutral 
component in each of these last two bidoublets, since 
their $B-L$ equals $\pm 2$, so the final mass matrix for 
the Higgs doublets is $4\times 4$. Only one eigenvalue of 
this matrix needs to be zero, and this can be achieved by 
fine-tuning. Each of the two Higgs doublets of the MSSM is 
thus a linear combination of 4 doublets, each of which has 
in general a vev of order ${\cal O}(M_W)$:

\begin{eqnarray}
\langle (2,2,1)_{10}\rangle&\approx&
\langle (2,2,15)_{\overline{126}}\rangle\approx
M_W\approx\\
\langle (2,2,15)_{126}\rangle&\approx&
\langle (2,2,10)_{210}\rangle\approx
\langle (2,2,\overline{10})_{210}\rangle\;.\nonumber
\end{eqnarray}

This mixing is nothing else than the susy version of 
\cite{Lazarides:1980nt,Babu:1992ia}.

(4) Due to all these bidoublet vevs, a SU(2)$_L$ triplet will 
also get a tiny but nonzero vev. Applying the susy constraint 
$F_{(3,1,10)_{126}}=0$ to 

\begin{equation}
W=M_{126}(3,1,10)_{126}(3,1,\overline{10})_{\overline{126}}+
(2,2,1)_{10}(2,2,\overline{10})_{210}(3,1,10)_{126}+...
\end{equation}

\noindent
one immediately gets

\begin{equation}
\langle (3,1,\overline{10})_{\overline{126}}\approx 
{\langle (2,2,1)_{10}\rangle 
\langle (2,2,\overline{10})_{210}\rangle\over
M_{126}}\approx {M_W^2\over M_{GUT}}\ne 0\;.
\end{equation}

This effect is just the susy version of 
\cite{Magg:1980ut}.

(5) Since $\nu$ lives in $(2,1,4)_{16}$, the second term in 
(\ref{wy}) gives among others also a term 
$(3,1,\overline{10})_{\overline{126}}(2,1,4)_{16}Y_{126}(2,1,4)_{16}$, 
which contributes to the light neutrino mass. So all together one gets 
for the light neutrino mass ($c$ is a 
model dependent dimensionless parameter)

\begin{equation}
\label{mn}
M_N=-M_{\nu_D}^TM_{\nu_R}^{-1}M_{\nu_D}+c{M_W^2\over M_{GUT}}Y_{126}\;.
\end{equation}

The first term is called the type I (or canonical) see-saw and is 
mediated by the SU(2)$_L$ singlet $\nu^c$, while the second is the 
type II (or non-canonical) see-saw, and is mediated by the SU(2)$_L$ 
triplet.

Equations (\ref{mu}), (\ref{md}), (\ref{mnd}), (\ref{me}), (\ref{mnr}) 
and (\ref{mn}) are all we need in the fit of known fermion masses and 
mixings and predictions of the unknown ones. A possible procedure is first 
to trade the matrices $Y_{10}$ and $Y_{126}$ for $M_U$ and $M_D$. The 
remaining freedom in $M_U$ and $M_D$ is still enough to fit $M_E$. 
But then some predictions in the neutrino sector are possible. For 
this sector we need to reproduce the experimental results 
$(\theta_l)_{12,23}\gg(\theta_q)_{12,23}$ and $(\theta_l)_{13}$ small. 
The degree of predictivity of the model however depends on the assumptions  
regarding the see-saw and on the CP phases.  

The first approach was to consider models in which type I dominates. 
It was shown that such models predict a small atmospheric neutrino 
mixing angle $\theta_{atm}=(\theta_l)_{23}$ if the CP phases are 
assumend to be small \cite{Babu:1992ia,Lee:1994je}. On the other hand, 
a large atmospheric neutrino mixing angle can be also large, if one 
allows for arbitrary CP phases and fine-tune them appropriately 
\cite{Matsuda:2001bg}.

A completely different picture emerges if one assumes that type II 
see-saw dominates. In this case even without CP violation one 
can naturally have a large atmospheric neutrino mixing angle, as 
has been first emphasized for the approximate case of second and 
third generations only in \cite{Bajc:2001fe,Bajc:2002iw}. In the three 
generation case the same result has been confirmed \cite{Goh:2003sy}. 
On top of this, a large solar neutrino mixing angle and a 
prediction of $U_{e3}\approx 0.15\pm 0.01$ (close to the 
upper experimental limit) have been obtained \cite{Goh:2003sy}.
Even allowing for general CP violation does not invalidate the 
above results: although the error bars are larger, the general 
picture of large atmospheric and solar neutrino mixing angles 
and small $U_{e3}$ still remains valid \cite{Goh:2003hf}.

It is possible to understand why type II see-saw gives so 
naturally a large atmospheric mixing angle. In type II the 
light neutrino mass matrix (\ref{mn}) is proportional to $Y_{126}$. 
From (\ref{md}) and (\ref{me}) one can easily find out, that 
$Y_{126}\propto M_D-M_E$, from which one gets \cite{Brahmachari:1997cq}

\begin{equation}
M_N\propto M_D-M_E\;.
\end{equation}

As a warm-up let us take the approximations of just 
(a) two generations, the second and the third, 
(b) neglect the masses of the second generation with 
respect of the third ($m_{s,\mu}\ll m_{b,\tau}$) and 
(c) assume that $M_D$ and $M_E$ has small mixings (this amounts 
to say, that in the basis of diagonal charged lepton mass, the smallness 
of the $({\theta_q})_{23}=\theta_{cb}$ is not caused by accidental 
cancellation of two large numbers). In this approximate set-up one gets 

\begin{equation}
M_N\propto
\pmatrix{
   0
&  0
\cr   
   0
&  m_b-m_\tau
\cr   }\;.
\end{equation}

This is, in type II see-saw there is a correspondence between 
the large atmospheric angle and $b-\tau$ unification \cite{Bajc:2002iw}.

Remember here that $b-\tau$ Yukawa unification is no more automatic, 
since we have also $\overline{126}_H$ Higgs on top of the usual $10_H$. 
It is however quite well satisfied phenomenologically.

One can do better: still take $m_{s,\mu}\approx 0$, but allow a 
nonzero quark mixing. In this case the atmospheric mixing angle is

\begin{equation}
\tan{2\theta_{atm}}={\sin{2\theta_{cb}}\over 
2\sin^2{\theta_{cb}}-{m_b-m_\tau\over m_b}}\;.
\end{equation}

Since $\theta_{cb}\approx {\cal O}(10^{-2})$, one again finds out 
the correlation between the large atmospheric mixing angle and 
$b-\tau$ unification at the GUT scale. 

The result can be confirmed of course also for finite $m_{s,\mu,c}$, 
although not in a so simple and elegant way.

Of course there are many other models that predict and/or explain a 
large atmospheric mixing angle (for a recent review see for example 
\cite{King:2003jb}). What is surprising here is, however, that no 
other symmetry except the gauge SO(10) is needed whatsoever.

\section{The minimal model}

As we saw in the previos section, one can correctly fit the known 
masses and mixings, get some understanding of the light neutrino 
mass matrix, and obtain some new predictions. What we would like 
to show here is that the model considered above has less number 
of model parameters than any other GUT, and can be then called 
the minimal realistic supersymmetric grand unified theory 
(even more minimal than SU(5)!) \cite{Aulakh:2003kg}. 

The Higgs sector described by (\ref{wh}) and (\ref{dwh}) contains 
$10$ real parameters ($7$ complex parameters minus four phase 
redefinitions due to the four complex Higgs multiplets involved). 
The Yukawa sector (\ref{wy}) has two complex symmetric matrices, 
one of which can be always made diagonal and real by a unitary 
transformation of $16$ in generation space. So what remains are 
$15$ real parameters. There is on top of this also the gauge 
coupling, so all together 26 real parameters in the supersymmetric 
sector of renormalizable SO(10) GUT with three copies of matter 16 and 
Higgses in the representations $10_H$, $126_H$, $\overline{126}_H$ and 
$210_H$. We will not consider the susy breaking sector, since this is 
present in all supersymmetric theories, GUTs or not.

Before comparing with other GUTs, for example SU(5), let us count 
the number of model parameters in MSSM. There are 6 quark masses, 
3 quark mixing angles, 1 quark CP phase, 6 lepton masses, 3 lepton 
mixing angles and 3 lepton CP phases (assuming Majorana neutrinos). 
On top of this, there are 3 gauge couplings and the real $\mu$ parameter. 
Thus, all together, again 26 real parameters. They are however distributed 
differently, so that in the Yukawa sector there are only 15 parameters 
in our minimal SO(10) GUT, which has to fit 22 MSSM (at least in principle) 
measurable low-energy parameters. Although in this fitting also few 
vevs that contain parameters from the Higgs and susy breaking sector 
play a role, the minimal SO(10) is nevertheless predictive.

One can play with other SO(10) models: the renormalizable ones need 
more representations and thus have more invariants, while the 
nonrenormalizable ones (those that use $16_H$ instead of $126_H$) have 
a huge number of invariants, some of which must be very small due to 
R-symmetry constraints. Of course, with some extra discrete, global or 
local symmetry, one can forbid these unpleasant and dangerous terms, 
remaining even with a small number of parameters, but as we said, this 
is not allowed in our scheme, in which we want to obtain as much 
information as possible just from GUT gauge symmetry (and 
renormalizability).

The simplicity of the minimal renormalizable supersymmetric SU(5) 
looks as if the number of parameters here could be smaller than in 
our previous example. What however gives a large number of parameters is 
the fact, that SU(5) is not particularly suitable for the neutrino sector. 
In fact, one can play and find out, that the minimal SU(5) with nonzero 
neutrino masses is obtained adding the two index symmetric $15_H$ and 
$\overline{15}_H$, and the number of model parameters comes out to be 
39, i.e. much more than in the minimal SO(10). 

\section{Conclusion}

Before talking about flavour symmetries it is important first to know, 
what we can learn from just pure GUTs. The minimal GUT is a SO(10) 
gauge theory with representations $10_H$, $126_H$, $\overline{126}_H$, 
$210_H$ and three generations of $16$. Such a realistic theory is 
renormalizable and no extra symmetries are needed. It can fit the 
fermion masses and mixings, and can give an interesting relation 
between $b-\tau$ Yukawa unification and large atmospheric mixing angle. 
It has a testable prediction for $U_{e3}$. Due to the large 
representations involved, it is not asymptotically free, which means 
that it predicts some new physics below $M_{Pl}$. 

There are many virtues of this minimal GUT. As in any SO(10) all 
fermions of one generation are in the same representation and 
the right-handed neutrino is included automatically, thus 
explaining the tiny neutrino masses by the see-saw mechanism. 
Employing $126_H$ instead of $16_H$ to break $B-L$ maintains 
R-symmetry exact at all energies. It is economical, it employs 
the minimal number of multiplets and parameters, and thus it is 
maximally predictive. It gives a good fit to available data and gives 
a framework to better understand the differences between the mixings in 
the quark and lepton sectors.

There are of course also some drawbacks. First, in order to maintain 
predictivity, one must believe in the principle of renormalizability, 
although the suppressing parameter in the expansion $M_{GUT}/M_{Planck}$ 
is not that small. Of course, in supersymmetry these terms can be small 
and stable, but this choice is not natural in the 't Hooft sense. 
Second, the DT splitting problem is here, and 
attempts to solve it require more fields \cite{Lee:1993jw}. 
Finally, usually it is said that 
$126$ dimensional representations are not easy to get from superstring 
theories, although we are probably far from a no-go theorem. 

There are many open questions to study in the context of the minimal 
SO(10), let me mention just few of them. 
First, proton decay: although it is generically dangerous, it is 
probably still possible to fit the data with some fine-tuning of 
the model parameters as well as of soft susy breaking terms. An 
interesting question is whether the model is capable of telling us 
which type of see-saw dominates. If it is type I or mixed, can it 
still give some testable prediction for $U_{e3}$? Also, gauge 
coupling unification should be tested in some way, although large 
threshold corrections could be nasty \cite{Dixit:1989ff}. And finally, 
is there some hope to solve in this context or minimal (but still 
predictive) extensions the doublet-triplet splitting problem? 

\section*{Acknowledgements}
It is a pleasure to thank the organizers for the well organized and 
stimulating conference. I am grateful to Charan Aulakh, 
Pavel Fileviez Perez, Alejandra Melfo, Goran Senjanovi\' c and Francesco 
Vissani for a fruitful collaboration. 
I thank Goran Senjanovi\' c also for carefully reading the 
manuscript and giving several useful advises and suggestions. 
This work has been supported by the Ministry of 
Education, Science, and Sport of the Republic of Slovenia.


\begin{thebibliography}{99}

\bibitem{Dimopoulos:1981yj}
S.~Dimopoulos, S.~Raby and F.~Wilczek,
Phys.\ Rev.\ D {\bf 24} (1981) 1681; 
L.~E.~Iba\~ nez and G.~G.~Ross,
Phys.\ Lett.\ B {\bf 105} (1981) 439; 
M.~B.~Einhorn and D.~R.~Jones,
Nucl.\ Phys.\ B {\bf 196} (1982) 475; 
W.~J.~Marciano and G.~Senjanovi\' c,
Phys.\ Rev.\ D {\bf 25} (1982) 3092.

\bibitem{Inoue:1982pi}
K.~Inoue, A.~Kakuto, H.~Komatsu and 
S.~Takeshita,
Prog.\ Theor.\ Phys.\  {\bf 68} (1982) 927
[Erratum-ibid.\  {\bf 70} (1983) 330]; 
L.~Alvarez-Gaume, J.~Polchinski and M.~B.~Wise,
Nucl.\ Phys.\ B {\bf 221} (1983) 495.

\bibitem{Lazarides:1980nt}
G.~Lazarides, Q.~Shafi and C.~Wetterich,
Nucl.\ Phys.\ B {\bf 181} (1981) 287.

\bibitem{Babu:1992ia}
K.~S.~Babu and R.~N.~Mohapatra,
Phys.\ Rev.\ Lett.\  {\bf 70} (1993) 2845
[arXiv:hep-ph/9209215].

\bibitem{Aulakh:1997ba}
C.~S.~Aulakh, K.~Benakli and G.~Senjanovi\' c,
Phys.\ Rev.\ Lett.\  {\bf 79} (1997) 2188
[arXiv:hep-ph/9703434]; 
C.~S.~Aulakh, A.~Melfo and G.~Senjanovi\' c,
Phys.\ Rev.\ D {\bf 57} (1998) 4174
[arXiv:hep-ph/9707256]; 
C.~S.~Aulakh, A.~Melfo, A.~Ra\v sin and G.~Senjanovi\' c,
Phys.\ Lett.\ B {\bf 459} (1999) 557
[arXiv:hep-ph/9902409].

\bibitem{Hisano:1992jj}
J.~Hisano, H.~Murayama and T.~Yanagida,
Nucl.\ Phys.\ B {\bf 402} (1993) 46
[arXiv:hep-ph/9207279]; 
T.~Goto and T.~Nihei,
Phys.\ Rev.\ D {\bf 59} (1999) 115009
[arXiv:hep-ph/9808255]; 
H.~Murayama and A.~Pierce,
Phys.\ Rev.\ D {\bf 65} (2002) 055009
[arXiv:hep-ph/0108104].

\bibitem{Bajc:2002bv}
B.~Bajc, P.~F.~Perez and G.~Senjanovi\' c,
Phys.\ Rev.\ D {\bf 66} (2002) 075005
[arXiv:hep-ph/0204311] 
and 
arXiv:hep-ph/0210374.

\bibitem{Masiero:1982fe}
A.~Masiero, D.~V.~Nanopoulos, K.~Tamvakis and T.~Yanagida,
Phys.\ Lett.\ B {\bf 115} (1982) 380.

\bibitem{Dimopoulos:xm}
S.~Dimopoulos and F.~Wilczek,
Print-81-0600 (SANTA BARBARA); 
K.~S.~Babu and S.~M.~Barr,
Phys.\ Rev.\ D {\bf 48} (1993) 5354
[arXiv:\-hep-\-ph/9306242] 
and 
Phys.\ Rev.\ D {\bf 65} (2002) 095009
[arXiv:hep-ph/0201130].

\bibitem{Berezhiani:1995sb}
see for example Z.~Berezhiani, C.~Csaki and L.~Randall,
Nucl.\ Phys.\ B {\bf 444} (1995) 61
[arXiv:hep-ph/9501336], and references therein.

\bibitem{Barbieri:1994jq}
R.~Barbieri, G.~R.~Dvali and A.~Strumia,
Phys.\ Lett.\ B {\bf 333} (1994) 79
[arXiv:hep-ph/9404278]; 
S.~M.~Barr,
Phys.\ Rev.\ D {\bf 55} (1997) 6775
[arXiv:hep-ph/9607359]; 
E.~Witten,
arXiv:hep-ph/0201018; 
M.~Dine, Y.~Nir and Y.~Shadmi,
Phys.\ Rev.\ D {\bf 66} (2002) 115001
[arXiv:hep-ph/0206268].

\bibitem{Bachas:1995yt}
C.~Bachas, C.~Fabre and T.~Yanagida,
Phys.\ Lett.\ B {\bf 370} (1996) 49
[arXiv:hep-th/9510094]; 
J.~L.~Chkareuli and I.~G.~Gogoladze,
Phys.\ Rev.\ D {\bf 58} (1998) 055011
[arXiv:hep-ph/9803335].

\bibitem{Dixit:1989ff}
V.~V.~Dixit and M.~Sher,
Phys.\ Rev.\ D {\bf 40} (1989) 3765.

\bibitem{Dvali:1992hc}
G.~R.~Dvali,
Phys.\ Lett.\ B {\bf 287} (1992) 101; 
P.~Nath,
Phys.\ Rev.\ Lett.\  {\bf 76} (1996) 2218
[arXiv:hep-ph/9512415]; 
P.~Nath,
Phys.\ Lett.\ B {\bf 381} (1996) 147
[arXiv:hep-ph/9602337]; 
V.~Lucas and S.~Raby,
Phys.\ Rev.\ D {\bf 55} (1997) 6986
[arXiv:hep-ph/9610293]; 
Z.~Berezhiani, Z.~Tavartkiladze and M.~Vysotsky,
arXiv:hep-ph/\-9809301; 
K.~Turzynski,
JHEP {\bf 0210} (2002) 044
[arXiv:hep-ph/0110282]; 
D.~Emmanuel-Costa and S.~Wiesenfeldt,
Nucl.\ Phys.\ B {\bf 661} (2003) 62
[arXiv:hep-ph/0302272].
S.~Rakshit, G.~Raz, S.~Roy and Y.~Shadmi,
arXiv:hep-ph/0309318.

\bibitem{Friedmann:2002ty}
T.~Friedmann and E.~Witten,
arXiv:hep-th/0211269; 
I.~R.~Klebanov and E.~Witten,
Nucl.\ Phys.\ B {\bf 664} (2003) 3
[arXiv:hep-th/0304079].

\bibitem{Mohapatra:yj}
R.~N.~Mohapatra,
Phys.\ Rev.\ Lett.\  {\bf 43} (1979) 893.

\bibitem{Nandi:1982ew}
S.~Nandi, A.~Stern and E.~C.~G.~Sudarshan,
Phys.\ Lett.\ B {\bf 113} (1982) 165; 
V.~S.~Berezinsky and A.~Y.~Smirnov,
Phys.\ Lett.\ B {\bf 140} (1984) 49.

\bibitem{Aulakh:2000sn}
C.~S.~Aulakh, B.~Bajc, A.~Melfo, A.~Ra\v sin and G.~Senjanovi\' c,
Nucl.\ Phys.\ B {\bf 597} (2001) 89
[arXiv:hep-ph/0004031].

\bibitem{Rubakov:fp}
V.~A.~Rubakov,
Nucl.\ Phys.\ B {\bf 203} (1982) 311; 
C.~G.~Callan,
Phys.\ Rev.\ D {\bf 25} (1982) 2141 
and 
Phys.\ Rev.\ D {\bf 26} (1982) 2058.

\bibitem{Freese:1983hz}
K.~Freese, M.~S.~Turner and D.~N.~Schramm,
Phys.\ Rev.\ Lett.\  {\bf 51} (1983) 1625.

\bibitem{Giacomelli:2003yu}
G.~Giacomelli and L.~Patrizii,
arXiv:hep-ex/0302011.

\bibitem{Melfo:2003xi}
A.~Melfo and G.~Senjanovi\' c,
Phys.\ Rev.\ D {\bf 68} (2003) 035013
[arXiv:hep-ph/0302216], and references therein.

\bibitem{Barbieri:1994pv}
R.~Barbieri and L.~J.~Hall,
Phys.\ Lett.\ B {\bf 338} (1994) 212
[arXiv:hep-ph/9408406].

\bibitem{Masina:2003iz}
I.~Masina and C.~Savoy,
arXiv:hep-ph/0309067.

\bibitem{Babu:2001qr}
K.~S.~Babu and R.~N.~Mohapatra,
Phys.\ Lett.\ B {\bf 518} (2001) 269
[arXiv:hep-ph/0108089].

\bibitem{Wilczek:1981iz}
F.~Wilczek and A.~Zee,
Phys.\ Rev.\ D {\bf 25} (1982) 553.

\bibitem{Albright:2000sz}
see for example 
C.~H.~Albright and S.~M.~Barr,
Phys.\ Rev.\ Lett.\  {\bf 85} (2000) 244
[arXiv:hep-ph/0002155]; 
K.~S.~Babu, J.~C.~Pati and F.~Wilczek,
Nucl.\ Phys.\ B {\bf 566} (2000) 33
[arXiv:hep-ph/9812538]; 
T.~Bla\v zek, S.~Raby and K.~Tobe,
Phys.\ Rev.\ D {\bf 60} (1999) 113001
[arXiv:hep-ph/9903340].

\bibitem{Chen:2003zv}
M.~C.~Chen and K.~T.~Mahanthappa,
arXiv:hep-ph/0305088.

\bibitem{Mohapatra:su}
R.~N.~Mohapatra,
Phys.\ Rev.\ D {\bf 34} (1986) 3457.

\bibitem{Clark:ai}
T.~E.~Clark, T.~K.~Kuo and N.~Nakagawa,
Phys.\ Lett.\ B {\bf 115} (1982) 26; 
D.~G.~Lee,
Phys.\ Rev.\ D {\bf 49} (1994) 1417.

\bibitem{Aulakh:1982sw}
C.~S.~Aulakh and R.~N.~Mohapatra,
Phys.\ Rev.\ D {\bf 28} (1983) 217.

\bibitem{Aulakh:2003kg}
C.~S.~Aulakh, B.~Bajc, A.~Melfo, G.~Senjanovi\' c and F.~Vissani,
arXiv:hep-ph/0306242.

\bibitem{Mohapatra:1979ia}
M.~Gell-Mann, P.~Ramond and R.~Slansky, proceedings of the
Supergravity Stony Brook Workshop, New York, 1979, eds. P.
Van Niewenhuizen and D. Freeman (North-Holland, Amsterdam);
T.~Yanagida, proceedings of the Workshop on Unified Theories and
Baryon Number in the Universe, Tsukuba, Japan 1979 (edited by
A. Sawada and A. Sugamoto, KEK Report No. 79-18, Tsukuba);
R.~N.~Mohapatra and G.~Senjanovi\' c,
Phys.\ Rev.\ Lett.\  {\bf 44} (1980) 912.

\bibitem{Magg:1980ut}
M.~Magg and C.~Wetterich,
Phys.\ Lett.\ B {\bf 94} (1980) 61; 
R.~N.~Mohapatra and G.~Senjanovi\' c,
Phys.\ Rev.\ D {\bf 23} (1981) 165.

\bibitem{Lee:1994je}
D.~G.~Lee and R.~N.~Mohapatra,
Phys.\ Rev.\ D {\bf 51} (1995) 1353
[arXiv:hep-ph/9406328].

\bibitem{Matsuda:2001bg}
K.~Matsuda, Y.~Koide, T.~Fukuyama and H.~Nishiura,
Phys.\ Rev.\ D {\bf 65} (2002) 033008
[Erratum-ibid.\ D {\bf 65} (2002) 079904]
[arXiv:hep-ph/0108202]; 
T.~Fukuyama and N.~Okada,
JHEP {\bf 0211} (2002) 011
[arXiv:hep-ph/0205066].

\bibitem{Bajc:2001fe}
B.~Bajc, G.~Senjanovi\' c and F.~Vissani,
arXiv:hep-ph/0110310.

\bibitem{Bajc:2002iw}
B.~Bajc, G.~Senjanovi\' c and F.~Vissani,
Phys.\ Rev.\ Lett.\  {\bf 90} (2003) 051802
[arXiv:hep-ph/0210207].

\bibitem{Goh:2003sy}
H.~S.~Goh, R.~N.~Mohapatra and S.~P.~Ng,
Phys.\ Lett.\ B {\bf 570} (2003) 215
[arXiv:hep-ph/0303055].

\bibitem{Goh:2003hf}
H.~S.~Goh, R.~N.~Mohapatra and S.~P.~Ng,
arXiv:hep-ph/0308197.

\bibitem{Brahmachari:1997cq}
B.~Brahmachari and R.~N.~Mohapatra,
Phys.\ Rev.\ D {\bf 58} (1998) 015001
[arXiv:hep-ph/9710371].

\bibitem{King:2003jb}
S.~F.~King,
arXiv:hep-ph/0310204.

\bibitem{Lee:1993jw}
D.~G.~Lee and R.~N.~Mohapatra,
Phys.\ Lett.\ B {\bf 324} (1994) 376
[arXiv:hep-ph/9310371].

\end{thebibliography}
\end{document}